\documentstyle[a4,11pt,amsmath,amssymb,latexsym,amssymb,epsfig,twoside,citesort]{article}
\def \d{{\mathrm{d}}}

\def \pd{\partial}

\def \tl#1{\overset{\kern 1pt\circ}{#1}}
\def \TL#1{\overset{\kern -3pt \circ}{#1}}
\def \TLL#1{\overset{\kern -7pt \circ}{#1}}

\begin{document}
\title{{\bf Screw dislocations in the field theory of elastoplasticity
	}}
\author{Markus Lazar\footnote{Present address: 
	Max-Planck-Institute for Mathematics in the Sciences,
        Inselstr. 22-26, D-04103 Leipzig, Germany.}
	\\
	Institute for Theoretical Physics, University of Leipzig \\
        Augustusplatz 10, D-04109 Leipzig, Germany \\
	E-mail: lazar@mis.mpg.de
	}
\date{August 9, 2002}    
\maketitle

\begin{abstract}
A (microscopic) static elastoplastic field theory of dislocations with 
moment and force stresses is considered. 
The relationship between the moment stress and the Nye
tensor is used for the dislocation Lagrangian.
We discuss the stress field of an infinitely long screw dislocation in 
a cylinder, a dipole of screw dislocations and a coaxial screw dislocation
in a finite cylinder.
The stress fields have no singularities in the dislocation core and 
they are modified in the core due to the presence of localized 
moment stress. 
Additionally, we calculated the elastoplastic energies for the screw dislocation
in a cylinder and the coaxial screw dislocation.
For the coaxial screw dislocation we find a modified formula
for the so-called Eshelby twist which 
depends on a specific intrinsic material length.\\

\end{abstract}
\vspace*{2mm}

\section{Introduction}
The traditional description of the elastic field produced by a dislocation is based 
on the classical theory of linear elasticity~\cite{Read,Cottrell,Nabarro,HL}. 
This approach works for the 
strain and stress field far from the core quite well. 
However, the 
components of these fields are singular at the dislocation line and
this theory, often applied to practical problems, misses the important feature
of plasticity.
This is unfortunate since the dislocation core is the most important region
in metallic plasticity.

Therefore, it is quite natural to think of dislocation theory as a 
field theory of 
elastoplasticity~\cite{Gairola81,Gairola93,Kroener93,Kroener96,lazar00,lazar01,malyshev00}. 
In this framework, it is possible to understand the dislocation density as
the elastoplastic field strength like the magnetic field strength in 
Maxwell's theory. 
The presence of dislocations leads to a specific response with the dimension 
of a moment stress, so that
the excitation or response quantity in elastoplasticity 
can be identified with a localized moment stress tensor.
In order to obtain a physical field theory of dislocations, 
we have to choose a constitutive law between dislocation density
and moment stress which is compatible with the symmetric force stress.
We use the constitutive law for the moment stress which is proportional to 
the Nye tensor.
A new material constant enters the constitutive law.
Additionally, it is possible to define a characteristic internal material 
length in this theory by means of the new material constant.

On the other hand, there are other non-standard continuum models of 
dislocations as, e.g., the nonlocal continuum 
model~\cite{eringen83,eringen85,eringen87,eringen90} 
and strain gradient elasticity~\cite{AA92,Aifantis94,GA96,GA99,Gutkin00,Fleck94}.
In these approaches also a characteristic length (gradient coefficient or
nonlocality parameter) appears.
In the case of a straight screw dislocation in an infinite medium, 
the force stresses in nonlocal elasticity~\cite{eringen83,eringen85,eringen87,eringen90} 
and in gradient elasticity~\cite{GA99,Gutkin00}
agree with the gauge theoretical ones~\cite{lazar01,malyshev00,VS88,Edelen96}. 
These solutions have no singularities at the dislocation line.

The plan of the paper is as follows. In Section~\ref{plastic}, the linear 
elastoplastic field theory of dislocations~\cite{lazar00,lazar01} is 
recalled. In this paper we use the widespread 
formalism of tensor analysis because we believe that this formulation 
of the elastoplastic field theory
is comprehensible and useful for 
material scientists, experimental physicists and engineers. 
We give the relation of this elastoplastic theory to Kr{\"o}ner's 
dislocation theory~\cite{kroener58,kroener81}. 
In Section~\ref{screw}, we investigate screw dislocations with different
boundary conditions in this framework.

\section{Linear elastoplastic field theory of dislocations}
\label{plastic}
In the elastoplastic field theory, the incompatible distortion tensor
\begin{align}
\label{dist-eff}
B_{ij}=\pd_j \xi_i+\phi_{ij},
\end{align}
plays a fundamental role~\cite{lazar00,lazar01}. Here $\xi_i$ is the mapping function from the 
ideal or reference configuration to the final or deformed configuration.
The first part of Eq.~(\ref{dist-eff}) describes the 
integrable part of the distortion,  
while $\phi_{ij}$ is, in general, the nonintegrable distortion 
which can be identified with the gauge potential of dislocations in the 
framework of the $T(3)$-gauge theory.
The form of Eq.~(\ref{dist-eff}) can be understood as a minimal 
replacement of the compatible distortion, $B_{ij}=\pd_j\xi_i$, in 
$T(3)$-gauge theory and is obtained from the translational
part of the generalized affine connection~\cite{lazar00,lazar01}. 
The distortion $B_{ij}$ describes the local distortion from the ideal or 
reference state to the dislocated state which is now anholonomic.

For simplicity, we will consider here a linear isotropic theory of dislocations. 
In this case, we do not distinguish between covariant and contravariant indices.
A nonlinear elastoplastic field theory of dislocations has been 
proposed in Refs.~\cite{lazar00,lazar01} (see also~\cite{malyshev00}).

If we use the weak field approximation (linearization) 
\begin{align}
\xi_i=x_i+u_i,\qquad  B_{ij}=\delta_{ij}+\beta_{ij},
\end{align}
where $u_i$ is the displacement field, 
the linear elastic strain tensor 
is given by means of the incompatible distortion tensor as
\begin{align}
E_{ij}\equiv\beta_{(ij)}=\frac{1}{2}\big(\pd_i u_j+\pd_ju_i+\phi_{ij}+\phi_{ji}\big).
\end{align}
The elastic strain energy of an isotropic material reads
\begin{align}
W=\frac{1}{2}\,\mu\left(\delta_{ik}\delta_{jl}+\delta_{il}\delta_{jk}
+\frac{2\nu}{1-2\nu}\,\delta_{ij}\delta_{kl}\right)E_{ij} E_{kl},
\end{align}
where $\mu$ is the shear modulus and $\nu$ Poisson's ratio.
The (symmetric) force stress is the response quantity (excitation)
with respect to the strain and is given by
\begin{align}
\sigma_{ij}=\frac{\pd W}{\pd E_{ij}}=
2\mu\left( E_{ij}+\frac{\nu}{1-2\nu}\,\delta_{ij} E_{kk}\right),\qquad
\sigma_{ij}=\sigma_{ji}.
\end{align}
The total strain $E^T_{ij}$ and distortion $\beta^T_{ij}$ are defined in 
terms of the total displacement field $u^T_i$. 
In the presence of dislocations, the total strain is not completely elastic,
but a part of it is stress-free or plastic~\cite{deWit} so that
\begin{align}
E^T_{ij}\equiv\pd_{(i}u^T_{j)}=E_{ij}+E^P_{ij},\quad
\beta^T_{ij}\equiv\pd_j u^T_i=\beta_{ij}+\beta^P_{ij}.
\label{total-strain}
\end{align}
The skew-symmetric part of the distortion tensor defines
the elastic rotation of a dislocation~\cite{deWit}
\begin{align}
\omega_i\equiv-\frac{1}{2}\,\epsilon_{ijk}\beta_{jk}.
\end{align}
The elastoplastic field strength is defined by means of the distortion
tensor (gauge potential)
as follows
\begin{align}
T_{ijk}:=\pd_j\beta_{ik}-\pd_k\beta_{ij}=
\pd_j\phi_{ik}-\pd_k\phi_{ij},\qquad T_{ijk}=-T_{ikj},
\end{align}
and
\begin{align}
T_{ijk}=-\pd_j\beta^P_{ik}+\pd_k\beta^P_{ij},
\end{align}
where $\beta^P_{ij}$ is the plastic distortion tensor
which may be identified with $-\phi_{ij}$.
The usual dislocation density $\alpha_{ij}$, which was originally introduced
by Nye~\cite{nye}, is recovered by
\begin{align}
\label{alpha-ij}
\alpha_{ij}:=\frac{1}{2}\,\epsilon_{jkl}T_{ikl}
            =\epsilon_{jkl}\pd_k \beta_{il} =-\epsilon_{jkl}\pd_k \beta^P_{il}.
\end{align}
In this approach the dislocation density may not assumed as a delta function.
It satisfies the translational Bianchi identity
\begin{align}
\label{bianchi}
\pd_j \alpha_{ij}=0,
\end{align}
which means that dislocations cannot end inside the body.
Note that the $\alpha_{ij}$ defined here is the transpose
of the dislocation density $\alpha_{ji}$ used by Kr{\"o}ner~\cite{kroener58}
(and $-\alpha_{ji}$ in~\cite{kroener81}).

We use the Lagrangian of dislocations in the following form~\cite{lazar00,lazar01}
\begin{align}
\label{L-core}
{\cal L}_{\rm disl}=
-\frac{1}{4}\, T_{ijk}\, H_{ijk},
\end{align}
with the moment stress $H_{ijk}$ 
as the elastoplastic excitation (response quantity) to the
dislocation density 
(torsion)\footnote{Here, the so-called Einstein-choice is used (see~\cite{lazar01}).}
\begin{align}
\label{moment1}
H_{ijk}=-2\frac{\pd {\cal L}_{\rm disl}}{\pd T_{ijk}}
=\frac{a_1}{2}\left(T_{ijk}- T_{jki} -T_{kij} 
-2\delta_{ij}T_{llk}-2\delta_{ik}T_{ljl}\right).
\end{align}
Here, $H_{ijk}=-H_{ikj}$.
The coefficient $a_1$ has the dimension of a force.
If we use the 
relations $H_{im}=\frac{1}{2}\epsilon_{jkm} H_{ijk}$ and
$T_{ijk}=\epsilon_{jkm} \alpha_{im}$, 
we find from Eq.~(\ref{moment1}) the relation between
the moment stress tensor $H_{ij}$ and the Nye~\cite{nye} 
tensor $\kappa_{ij}$ as
\begin{align}
\label{nye}
H_{ij}= a_1\left(\alpha_{ji}-\frac{1}{2}\,\delta_{ij}\alpha_{kk}\right)
        \equiv a_1\kappa_{ij}.
\end{align}
Hence, this moment stress tensor $H_{ij}$ is proportional to the Nye 
tensor (see also~\cite{hehl65}). 
The diagonal components $H_{ii}$ describe twisting and the 
nondiagonal components of $H_{ij}$ bending moments localized in the 
dislocation core region.
Alternatively, the dislocation Lagrangian~(\ref{L-core}) can be written as
\begin{align}
{\cal L}_{\rm disl}=
-\frac{a_1}{2}\, \alpha_{ij} \kappa_{ij}.
\end{align}

In the framework of elastoplastic field theory,
we require that the modified stress field of a dislocation 
has the following properties:
(i)~the stress field should have no singularity at $r=0$,
(ii)~the far field stress ought to be the stress field of a 
Volterra dislocation (background stress) $\tl\sigma_{ij}$ which 
satisfies the condition $\pd_j\tl\sigma_{ij}=0$. 
The boundary Lagrangian is given by a so-called
null Lagrangian~\cite{edelen88} and by using the minimal replacement 
$\pd_j u_i\rightarrow \beta_{ij}$: 
\begin{align}
W_{\rm bg}=\pd_j\big(\tl\sigma_{ij} u_i\big)
          =\big(\pd_j \tl\sigma_{ij}\big) u_i
            +\tl\sigma_{ij} \pd_j u_i
          \rightarrow\tl\sigma_{ij}\beta_{ij}.
\end{align} 

The Euler-Lagrange equations for ${\cal L}={\cal L}_{\rm disl}-W+W_{\rm bg}$ 
take the form
\begin{align}
\label{feq}
\pd_j\sigma_{ij}&=0& &\text{(force equilibrium)},\\
\label{meq}
\pd_k H_{ikj} &=\widehat\sigma_{ij}& &\text{(moment equilibrium)},
\end{align}
where the so-called effective stress tensor, 
\begin{align}
\widehat\sigma_{ij}:=\sigma_{ij}-\tl\sigma_{ij}, 
\end{align}
is the driving ``source'' for dislocations. 
The left hand side of Eq.~(\ref{meq}) reads explicitly 
\begin{align}
\label{Einstein}
\frac{1}{a_1}\,\pd_k H_{ikj}
&=-\frac{1}{a_1}\,\pd_k H_{(ij)k}\nonumber\\
&=\Delta E_{ij}-(\pd_j\pd_k E_{ik}+\pd_i\pd_k E_{kj})
+\delta_{ij}\pd_k\pd_l E_{kl}+\pd_i\pd_j E_{kk}
-\delta_{ij}\Delta E_{kk}\nonumber\\
&\equiv\left(\text{inc}\, \boldsymbol{E}\right)_{ij}
=-\epsilon_{ikl}\epsilon_{jmn}\pd_k \pd_{m} E_{l n},
\end{align}
and
\begin{align}
\label{Einstein-skew}
\pd_k H_{[ij]k}=0.
\end{align}
Consequently, the antisymmetric moment equilibrium or equilibrium of couple
stress (\ref{Einstein-skew}) is trivially satisfied in the Einstein choice
and, thus, no antisymmetric force stresses appear.
Additionally, we obtain the equation for the elastic strain $ E_{ij}$ as
\begin{align}
\label{fe-E}
a_1\big\{\Delta E_{ij}-(\pd_j\pd_k E_{ik}+\pd_i\pd_k E_{kj})
+\delta_{ij}\pd_k\pd_l E_{kl}+\pd_i\pd_j E_{kk}
-\delta_{ij}\Delta E_{kk}\big\}=\widehat\sigma_{ij}.
\end{align}
This is the incompatibility condition of the elastoplastic 
theory of dislocations.
By means of the inverse of Hooke's law
\begin{align}
 E_{ij}=\frac{1}{2\mu}\left(\sigma_{ij}-\frac{\nu}{1+\nu}\,\delta_{ij}\sigma_{kk}\right),
\end{align}
and the equilibrium condition $\pd_j\sigma_{ij}=0$, 
we obtain the field equation for the stress field
\begin{align}
\label{beltrami}
\Delta\sigma_{ij}+\frac{1}{1+\nu}\big(\pd_i\pd_j
-\delta_{ij}\Delta\big)\sigma_{kk}=\kappa^2\, \widehat\sigma_{ij},
\qquad \kappa^2=\frac{2\mu}{a_1}.
\end{align}
For $\widehat\sigma_{ij}=0$, this equation is the Beltrami equation.
The coefficient $\kappa$ has the dimension of a reciprocal length.
Now, we compare Eq.~(\ref{fe-E}) with Kr{\"o}ner's incompatibility 
condition~\cite{kroener58,kroener81}
\begin{align}
-\epsilon_{ikl}\epsilon_{jmn}\pd_k \pd_{m} E_{l n}=\eta_{ij},
\end{align}
where the so-called incompatibility tensor $\eta_{ij}$ is defined by
\begin{align}
\label{inc1}
\eta_{ij}:=-\frac{1}{2}\big(\epsilon_{ikl}\pd_k\alpha_{lj}+
                  \epsilon_{jkl}\pd_k\alpha_{li}\big).
\end{align}
Then we can identify
\begin{align}
\label{inc2}
\eta_{ij}\equiv\frac{1}{a_1}\,\widehat\sigma_{ij}.
\end{align}  
Therefore, Eq.~(\ref{fe-E}) represents the proper gauge theoretical 
formulation of Kr{\"o}ner's incompatibility equation. 
However, Edelen's field equation (see, e.g., Eq.~(9-8.7) in Ref.~\cite{edelen88})
\begin{align}
\Delta \beta_{ij}-\pd_j\pd_l \beta_{il}=
\frac{1}{a_1}\, \widehat\sigma_{ij},
\end{align}
cannot be interpreted in this way, in contrast to his claim. 

We may define the tensor
\begin{align}
\Theta_{ij}:=-\epsilon_{jkl}\pd_k \kappa_{il}
\end{align}
which can be identified with the linearized Einstein tensor in
a Riemannian space. Here, $\Theta_{ij}$ is determined by the dislocation 
density.
Using $H_{ijk}=\epsilon_{jkl} H_{il}$ and Eq.~(\ref{nye}),
the field equation (\ref{meq}) has the shape of
a linearized Einstein-type field equation according 
to (see also~\cite{lazar00,lazar01,malyshev00})
\begin{align}
a_1 \Theta_{ij}=\widehat\sigma_{ij}\qquad
{\text{with }}\ 
\pd_j\widehat\sigma_{ij}=0,\quad
\pd_j \Theta_{ij}=0.
\end{align}

\section{Screw dislocations in the elastoplastic field theory}
\label{screw}
\subsection{Screw dislocation in a cylinder}
We now consider an infinitely long screw dislocation whose dislocation line 
and Burgers vector 
coincide with the axis of an isotropic cylinder of radius $R$.
Due to the symmetry of the problem, we choose the Burgers vector and the 
dislocation line in $z$-direction: $b_x=b_y=0$, $b_z=b$.
Using the stress function ansatz~\cite{HL,kroener58,kroener81}, 
the elastic stress of a Volterra screw dislocation, 
in Cartesian coordinates, reads 
\begin{align}
\tl\sigma_{xz}=
\tl\sigma_{zx}=-\pd_y \Phi=-\frac{\mu b}{2\pi}\,\frac{y}{r^2},
\quad
\tl\sigma_{yz}=
\tl\sigma_{zy}=\pd_x\Phi=\frac{\mu b}{2\pi}\,\frac{x}{r^2},
\end{align}
or, in cylindrical coordinates,
\begin{align}
\label{T-bg}
\tl\sigma_{z\varphi}=
\tl\sigma_{\varphi z}=\pd_r\Phi=\frac{\mu b}{2\pi r},
\end{align}
where $r^2=x^2+y^2$.
Here, $\Phi$ is the well-known stress function of elastic torsion, sometimes 
called Prandtl's stress function. It is given by (see, e.g.,~\cite{kroener81})
\begin{align}
\Phi=\frac{\mu b}{2\pi}\, \ln r.
\end{align}
This is Green's function of the 2-dimensional potential equation
\begin{align}
\Delta \Phi=\mu b\, \delta(r).
\end{align}
Obviously, the ``classical'' stress fields are singular at the dislocation line.
For simplicity, we assume that the cylindrical surface $r=R$ is free 
of tractions, i.e.,
\begin{align}
\tl\sigma_{rr}=\tl\sigma_{r\varphi}=\tl\sigma_{\varphi r}=
\tl\sigma_{rz}=\tl\sigma_{zr}=0\qquad{\text{for}}\ r=R.
\end{align}
For the modified stresses we make the ansatz
\begin{align}
\sigma_{xz}=\sigma_{zx}=-\pd_y F,
\qquad
\sigma_{yz}=\sigma_{zy}=\pd_x F,
\end{align}
where $F$ is called the modified Prandtl stress function.
We turn to Eq.~(\ref{beltrami}) and use $\sigma_{kk}=0$.
In this way, the field equation for the force stress of a linear screw dislocation 
reduces to an inhomogeneous Helmholtz equation, 
\begin{align}
\label{stress-fe-screw}
\left(1-\kappa^{-2}\Delta\right)\sigma_{ij}=\tl\sigma_{ij},
\end{align} 
with the ``source'' term given in terms of the well-known solutions 
of classical dislocation theory for the same traction boundary-value
problem.
Thus, for the elastic strain fields,
\begin{align}
\left(1-\kappa^{-2}\Delta\right)E_{ij}=\tl E_{ij},
\end{align} 
where $\tl E_{ij}$ is the ``classical'' strain tensor.
Substituting the stress functions into (\ref{stress-fe-screw}), 
we get
\begin{align}
\left(1-\kappa^{-2}\Delta\right)F=\frac{\mu b}{2\pi}\, \ln r.
\end{align} 
The solution of the modified stress function of a screw dislocation
is given by
\begin{align}
F=\frac{\mu b}{2\pi}\Big\{\ln r +K_0(\kappa r)\Big\},
\end{align}
where $K_n$ is the modified Bessel function of the second kind and of 
order $n$.
Consequently, we find the force stresses
\begin{align}
\sigma_{xz}=\sigma_{zx}=-\frac{\mu b}{2\pi}\,\frac{y}{r^2}\Big\{1-\kappa r K_1(\kappa r)\Big\},
\quad
\sigma_{yz}=\sigma_{zy}=\frac{\mu b}{2\pi}\,\frac{x}{r^2}\Big\{1-\kappa r K_1(\kappa r)\Big\},
\end{align}
and in cylindrical coordinates
\begin{align}
\label{T-cyl}
\sigma_{z\varphi}=\sigma_{\varphi z}=\frac{\mu b}{2\pi}\,\frac{1}{r}\Big\{1-\kappa r K_1(\kappa r)\Big\}.
\end{align}
If we choose the coefficient $\kappa$ as ($a$ = lattice constant)
\begin{align}
\kappa^{-1}\approx 0.4 a,
\end{align}
the stress $\sigma_{zy}$ has its maximum $0.4\kappa\mu b/2\pi=\mu b/2\pi a$ at 
$x\simeq 1.1/\kappa=0.44 a$ whereas the minimum 
$-0.4\kappa\mu b/2\pi=-\mu b/2\pi a$ lies at $x\simeq -1.1/\kappa=-0.44 a$.
\begin{figure}[t]\unitlength1cm
\centerline{
\epsfxsize=9cm\epsffile{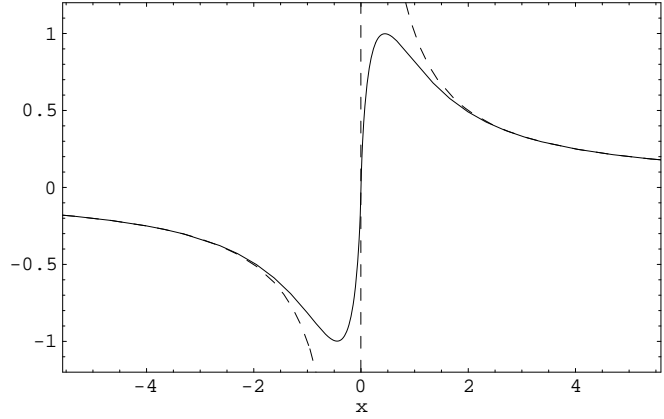}}
\caption{Modified stress field $\sigma_{yz}(x,0)/(\mu b/2\pi)$
(solid), with $\kappa^{-1}\simeq 0.4$, $a=1$, and classical stress (dashed) 
of a screw dislocation plotted over the $x$-coordinate.}
\label{fig:stress-screw}
\end{figure}
This is in agreement with the Peierls-Nabarro model. 
Note that the maximum stress may be identified with the 
theoretical shear strength.
The core radius $r_c$ is then given as $r_c\simeq 2.4 a$
(Figure~\ref{fig:stress-screw}).
This estimate of $\kappa$ is in conformity with
the result in Eringen's nonlocal elasticity theory. He
pointed out that for the choice of $\kappa^{-1}\approx 0.4 a$ 
one has an excellent fit between atomic dispersion curve and 
the nonlocal result~\cite{eringen83,eringen85,eringen87,eringen90}. 
However,
the factor $\kappa$ should be fitted by comparing predictions of the theory 
with experimental results. 
In general, the parameter $\kappa$ can be used to determine the width of a dislocation 
and the amplitude of the force stress.

Let us now calculate the distortion of a screw dislocation.
The distortion $\beta_{ij}$ is given in terms of the stress function
\begin{align}
\label{dist-ansatz}
\beta_{ij}=\frac{1}{2\mu}
\left(\begin{array}{ccc}
0 & 0 & -\pd_{y} F +2\mu\omega_1\\
0 & 0 & \pd_{x} F-2\mu\omega_2\\
-\pd_{y} F -2\mu\omega_1 &\pd_{x} F+2\mu\omega_2  & 0
\end{array}\right),
\end{align}
where $\omega_1$ and $\omega_2$ are used to express the antisymmetric part 
of the distortion.
Eventually, $\omega_1$ and $\omega_2$  are determined from the conditions
\begin{align}
\alpha_{xy}=T_{xzx}&=-\frac{1}{2\mu}\, \pd_x\big(2\mu\omega_1-\pd_y F\big)\equiv 0,\\
\alpha_{yx}=T_{yyz}&=-\frac{1}{2\mu}\, \pd_y\big(2\mu\omega_2-\pd_x F\big)\equiv 0.
\end{align}
One finds for the distortion tensor of the screw dislocation 
\begin{align}
\beta_{zx}=-\frac{b}{2\pi}\,\frac{y}{r^2}\Big\{1-\kappa r K_1(\kappa r)\Big\},
\quad
\beta_{zy}=\frac{b}{2\pi}\,\frac{x}{r^2}\Big\{1-\kappa r K_1(\kappa r)\Big\},
\end{align}
and for elastic rotation vector
\begin{align}
\omega_x\equiv\omega_2=\frac{b}{4\pi}\,\frac{x}{r^2}\Big\{1-\kappa r K_1(\kappa r)\Big\},
\quad
\omega_y\equiv\omega_1=\frac{b}{4\pi}\,\frac{y}{r^2}\Big\{1-\kappa r K_1(\kappa r)\Big\}.
\end{align}
The rotation vector is in agreement with the result calculated in the 
Cosserat theory~\cite{Nowacki73}.

By means of the distortion tensor it is possible to determine the 
effective Burgers vector as
\begin{align}
b_z(r)=\oint_\gamma\big(\beta_{zx}\d x+\beta_{zy}\d y\big)
      =b\big\{1-\kappa r K_1(\kappa r)\big\},
\end{align}
where $\gamma$ is the Burgers circuit.
This effective Burgers vector differs appreciably from the constant 
value $b$ in the core region from $r=0$ up to $r\simeq 6/\kappa$. Thus, it is 
suggestive to take $r_c\simeq 6/\kappa$ as the characteristic length 
(dislocation core radius). 
Outside this core region, the Burgers vector reaches its constant value. 
Accordingly, the classical and the elastoplastic solution coincide outside 
the core region.

If we use the decomposition~(\ref{total-strain}) of the elastic distortion 
into the total and the plastic distortion,
the total displacement field turns out to be
\begin{align}
\label{u-T}
u^T_z
=\frac{b}{2\pi}\,\varphi \Big\{1-\kappa r K_1(\kappa r)\Big\},\quad
\varphi=\arctan\frac{y}{x}.
\end{align}
Thus, the proper incompatible part of the total distortion 
is the plastic distortion which is confined in the 
dislocation core region 
\begin{align}
\label{plastic-dist}
\beta^P_{zx}
=\frac{b\kappa^2}{2\pi}\,x \varphi K_0(\kappa r),
\quad
\beta^P_{zy}
=\frac{b\kappa^2}{2\pi}\,y \varphi K_0(\kappa r).
\end{align}
It fulfills Eq.~(\ref{alpha-ij}).
When $\kappa^{-1}\rightarrow 0$, the plastic
distortion~(\ref{plastic-dist}) converts to 
$\beta^P_{zx}=b\,x\varphi\,\delta(r)$ and $\beta^P_{zy}=b\,y\varphi\,\delta(r)$.
In our framework,
Eq.~(\ref{u-T}) describes the ``atomic'' arrangement in 
the dislocation core region in contrast to classical elasticity which 
leads to an unphysical singularity.
The incompatible elastic and plastic strains read, respectively,
\begin{align}
E_{xz}=E_{zx}=-\frac{b}{4\pi}\,\frac{y}{r^2}\Big\{1-\kappa r K_1(\kappa r)\Big\},
\ \ 
E_{yz}=E_{zy}=\frac{b}{4\pi}\,\frac{x}{r^2}\Big\{1-\kappa r K_1(\kappa r)\Big\},
\end{align}
and 
\begin{align}
E^P_{xz}=E^P_{zx}=\frac{b\kappa^2}{4\pi}\, x\varphi K_0(\kappa r),\quad
E^P_{yz}=E^P_{zy}=\frac{b\kappa^2}{4\pi}\, y\varphi K_0(\kappa r).
\end{align}

Now we are able to calculate the dislocation density by means of the
distortion tensor. We obtain
\begin{align}
\label{alpha-zz}
\alpha_{zz}=T_{zxy}=-T_{zyx}
=\frac{1}{\mu}\, \Delta F=\frac{b\kappa^2}{2\pi}\, K_0(\kappa r).
\end{align}
Of course, this dislocation density satisfies the conditions 
(\ref{bianchi}), (\ref{inc1}) and (\ref{inc2}).
In the limit as $\kappa^{-1}\rightarrow 0$, the elastoplastical 
result~(\ref{alpha-zz}) converts to the classical dislocation density
$\alpha_{zz}=b\,\delta(r)$.
The localised moment stresses caused by the screw dislocation 
can be expressed in terms of the dislocation density as
\begin{align}
H_{zz}=a_1\kappa_{zz}=\frac{\mu}{\kappa^2}\,\alpha_{zz},\ \
H_{xx}=a_1\kappa_{xx}=-\frac{\mu}{\kappa^2}\,\alpha_{zz},\ \
H_{yy}=a_1\kappa_{yy}=-\frac{\mu}{\kappa^2}\,\alpha_{zz},
\end{align}
and 
\begin{align}
H_{kk}=-\frac{\mu}{\kappa^2}\,\alpha_{zz},\qquad \kappa_{kk}=-\frac{1}{2}\,\alpha_{zz}.
\end{align}
Accordingly, moment stresses of twisting-type occur at all positions where 
the dislocation density $\alpha_{zz}$ is non-vanishing.
When $\kappa^{-1}\rightarrow 0$, the moment stresses vanish. 

We next define the characteristic intrinsic material length by
\begin{align}
R_c:=\frac{1}{\kappa}=\sqrt{\frac{a_1}{2\mu}}.
\end{align}
$R_c$ measures the region over which the dislocation density (torsion),
the Nye tensor and the moment stress are appreciably different from zero.
It can be called the plastic penetration depth. 
Then $R_c$ is the internal length which measures the plastic region where 
$\beta^P_{ij}\neq 0$. In this region the local atomic configuration 
is fundamentally different from that of the defect-free part of the crystal. 
An important consequence is that the mechanical but also the electrical properties are 
different from those in the undefected crystal.

The strain energy density produced by the screw dislocation is
\begin{align}
W=\frac{1}{2}\, \sigma_{ij}E_{ij}
 =\frac{\mu b^2}{8\pi^2r^2}\,\big(1-\kappa r K_1(\kappa r)\big)^2.
\end{align}
We find the strain energy stored per unit length of the screw dislocation 
in the cylinder as
\begin{align}
\label{E_strain}
 E_{\rm strain}&=\int_{0}^R\!\d r\, r\int_0^{2\pi}\!\d\varphi\, W\nonumber\\
               &=\frac{\mu b^2}{4\pi} 
                \Big\{\ln r+2K_0(\kappa r)
               +\frac{\kappa^2 r^2}{2}\big(K_1(\kappa r)^2
                -K_0(\kappa r)K_2(\kappa r)\big)\Big\}\Big|_{0}^R, 
\end{align}
where $R$ is the outer cut-off radius.  
We use the limiting expressions for $r\rightarrow 0$,
\begin{align}
K_0(\kappa r)\approx-\gamma-\ln \frac{\kappa r}{2},\quad
K_1(\kappa r)\approx \frac{1}{\kappa r},\qquad
K_2(\kappa r)\approx -\frac{1}{2}+\frac{2}{(\kappa r)^2},
\end{align}
where $\gamma=0.57721566\ldots$ is the Euler constant.
The final result reads 
\begin{align}
\label{E_strain2}
 E_{\rm strain}=\frac{\mu b^2}{4\pi} 
                \Big\{\ln\frac{\kappa R}{2}+\gamma&-\frac{1}{2}
                +2K_0(\kappa R)\nonumber\\
                &+\frac{\kappa^2 R^2}{2}\big(K_1(\kappa R)^2
                -K_0(\kappa R)K_2(\kappa R)\big)\Big\}.
\end{align}
Thus, we obtain a strain energy expression which is not singular at the 
dislocation line and at $R$.

The dislocation core energy density produced by the screw dislocation 
is given by
\begin{align}
V_{\rm core}=\frac{a_1}{2}\, \alpha_{zz}\kappa_{zz}
            =\frac{\mu b^2\kappa^2}{8 \pi^2}\, K_0(\kappa r)^2.
\end{align}
The dislocation core energy per unit length of the screw dislocation is
\begin{align}
\label{E_core}
E_{\rm core}&=\int_0^R\!\d r\, r\int_0^{2\pi}\!\d\varphi\, V_{\rm core} \nonumber\\ 
            &=\frac{\mu b^2\kappa^2}{8\pi}\, 
                r^2\Big\{K_0(\kappa r)^2-K_1(\kappa r)^2\Big\}\Big|_0^R\nonumber\\
            &=\frac{\mu b^2}{8\pi}\Big\{1 +
                \kappa^2 R^2\big(K_0(\kappa R)^2-K_1(\kappa R)^2\big)\Big\}.            
\end{align}
Eventually, the total energy per unit length of the screw dislocation in
a cylinder is
given by
\begin{align}
E_{\rm screw}&=E_{\rm strain}+E_{\rm core}\nonumber\\
             &=\frac{\mu b^2}{4\pi} 
                \left\{\ln\frac{\kappa R}{2}+\gamma
                +2 K_0(\kappa R)-\kappa R\, K_0(\kappa R) K_1(\kappa R)\right\}.
\end{align}
It depends on the radius $R$ of the cylinder.

\subsection{A dipole and parallel screw dislocations}
Because the strain and stress fields in our approach 
have not any singularities, it is interesting to discuss the
stress fields of two (anti)-parallel screw dislocations ($A$) and 
($B$) near the dislocation cores. 
These two dislocations lie in the ($xy$)-plane at $x=0$, $y=0$ ($A$) and
$x=d$, $y=0$ ($B$) with the Burgers vector extending in $z$-direction.
Eventually, the stress fields of these two screw dislocations
are given by the superpositions of the corresponding fields of individual
ones as
\begin{align}
\sigma^{\|}_{yz}&=\frac{\mu}{2\pi}\left(
\frac{b_A x}{r^2_A}\Big\{1-\kappa r_A K_1(\kappa r_A)\Big\}
+\frac{b_B(x-d)}{r^2_B}\Big\{1-\kappa r_B K_1(\kappa r_B)\Big\}\right),\nonumber\\
\sigma^{\|}_{xz}&=-\frac{\mu} {2\pi}\left(
\frac{b_A y}{r^2_A}\Big\{1-\kappa r_A K_1(\kappa r_A)\Big\}
+\frac{b_B y}{r^2_B}\Big\{1-\kappa r_B K_1(\kappa r_B)\Big\}\right).
\end{align}
Here
$r^2_A=x^2+y^2$ and $r^2_B=(x-d)^2+y^2$.
For $b_A=b_B$ and $b_A= -b_B$, this configuration describes
two parallel screw dislocations and a dipole of screw dislocations, 
respectively.
\begin{figure}[t]
\centerline{
(a)
\epsfxsize=9cm\epsffile{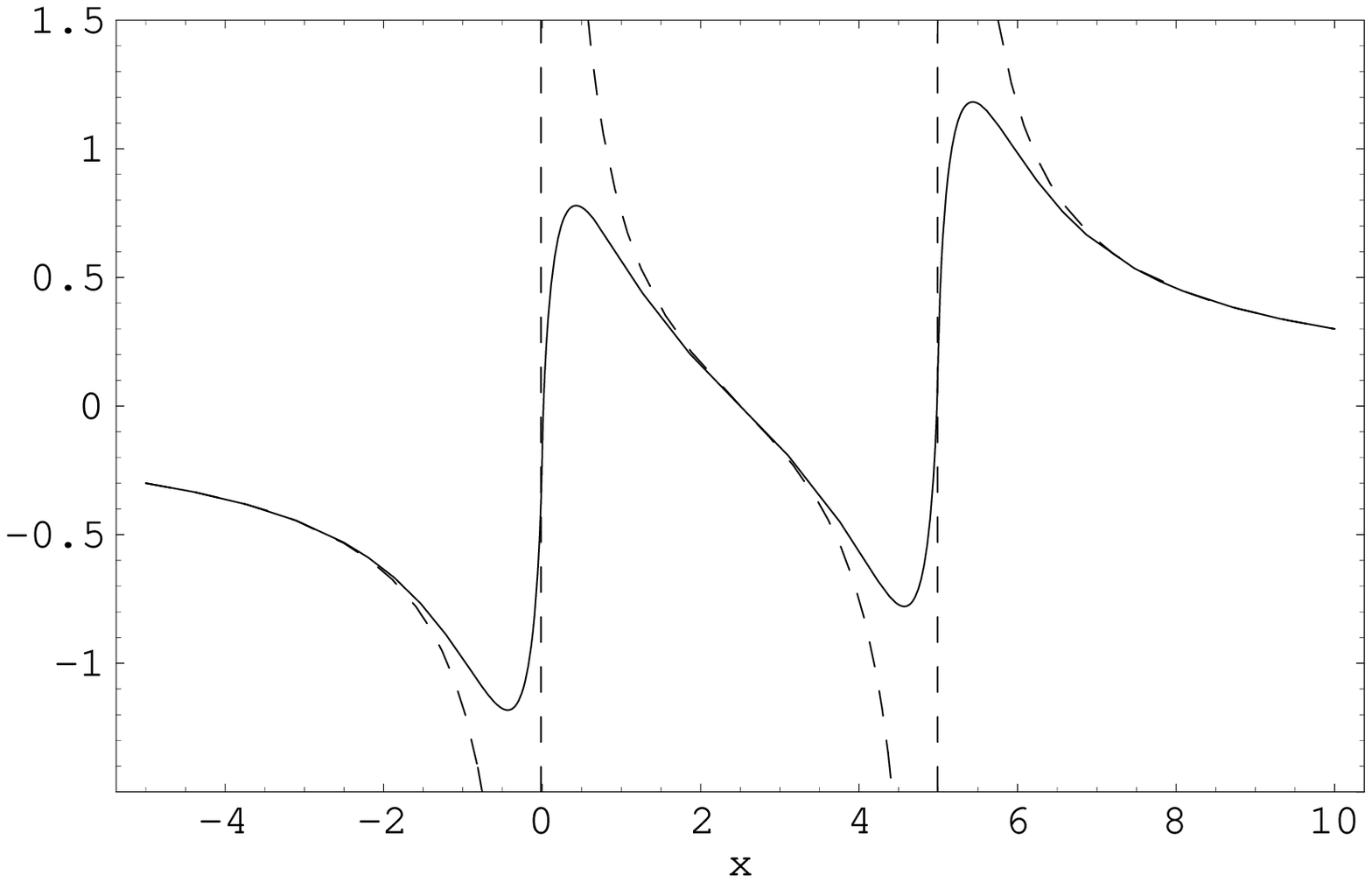}}
\vspace*{0.2cm}
\centerline{
(b)
\epsfxsize=9cm\epsffile{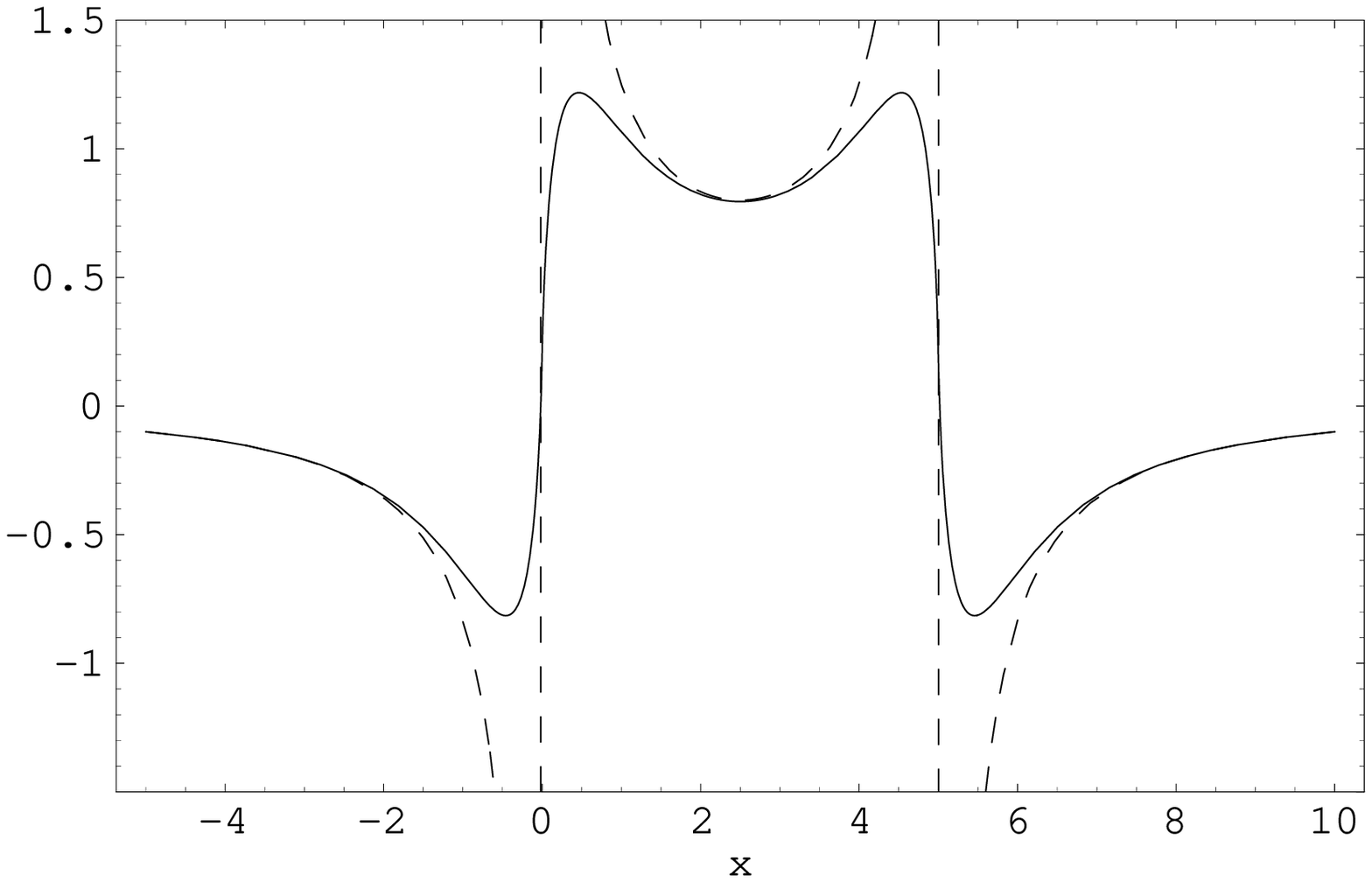}}
\caption{The stress component $\sigma^{\|}_{yz}(x,0)/(\mu b_A/2\pi)$, with $d=5$,
$\kappa^{-1}=0.4$ and $a=1$ (solid),
(a) of a pair of parallel screw dislocations,
(b) of a screw dislocation dipole. The dashed curves represent the classical
stress component.}
\label{fig:stress-dipole}
\end{figure}
Again, in contrast to the classical stress fields, the modified stress
fields dislpay a smooth shape and have no singularity 
(Figure~\ref{fig:stress-dipole}). 

We may discuss the stress field of a wall of screw dislocations.
A finite wall of (anti) parallel screw dislocations lying in the plane $x=0$ 
is obtained by a superposition of the individual screw dislocations.
The stress field of this wall is given by
\begin{align}
\sigma^{\|}_{yz}=\sum_{n=0}^{N}
\frac{\mu b_n}{2\pi}
\frac{x}{r^2_n}\Big\{1-\kappa r_n K_1(\kappa r_n)\Big\},\quad
\sigma^{\|}_{xz}=-\sum_{n=0}^{N}
\frac{\mu b_n} {2\pi}
\frac{y_n}{r^2_n}\Big\{1-\kappa r_n K_1(\kappa r_n)\Big\},
\end{align}
where $N+1$ is the number of the dislocations in the wall, $b_n$ is the 
Burgers vector of the individual dislocations,
$r^2_n=x^2+y_n^2$, $y_n=y-n d$ and $d$ is the spacing of the 
dislocations.

Because the stress fields have no singularities,
it is possible to study the behaviour and the interaction between the
screw dislocations in this framework.

\subsection{A coaxial screw dislocation in a cylinder}
Now we consider a screw dislocation in a cylinder of finite length in 
$z$-direction. Up to now, this problem 
was only discussed in classical 
elasticity (see~\cite{Cottrell,HL,mann,eshelby,teodosiu}).
We use the solution~(\ref{T-cyl}) which fulfills the boundary condition at
the cylindrical surface $r=R$. 
On the end of the surfaces the solution~(\ref{T-cyl}) 
produces the torque
\begin{align}
M_z&=\int_0^{2\pi}\!\d\varphi\int_0^R\!\d r\,r^2 \sigma_{z\varphi}\nonumber\\
   &=\frac{\mu b R^2}{2}\Big\{1-\frac{4}{\kappa^2 R^2}+2K_2(\kappa R)\Big\}
\end{align}
on the upper basis and $-M_z$ on the lower basis of the cylinder.
Thus, Eq.~(\ref{T-cyl}) is not the correct solution for a finite rod and it 
has to be modified. 
The torque is removed by adding an equal and opposite one by means of
an additional background condition
\begin{align}
&\tl u {}'_x=\tau yz,\qquad \tl u {}'_y=-\tau xz,\nonumber\\
&\tl\sigma {}'_{xz}=\tl\sigma {}'_{zx}=\tau \mu y,\qquad
\tl\sigma {}'_{yz}=\tl\sigma {}'_{zy}=-\tau \mu x,
\end{align}
with a coefficient $\tau$ which is the so-called Eshelby twist.
Then the correct boundary condition for the stress field is
\begin{align}
\tl\sigma_{xz}=\tl\sigma_{zx}=-\frac{\mu b}{2\pi }\frac{y}{r^2}+\tau \mu y,\qquad
\tl\sigma_{yz}=\tl\sigma_{zy}=\frac{\mu b}{2\pi }\frac{x}{r^2}-\tau \mu x.
\end{align}
Eventually we find from Eq.~(\ref{stress-fe-screw})
\begin{align}
&\sigma_{xz}=\sigma_{zx}=
-\frac{\mu b}{2\pi}\,\frac{y}{r^2}\Big\{1-\kappa r K_1(\kappa r)\Big\}
+\tau\mu y,\nonumber\\
&\sigma_{yz}=\sigma_{zy}=
\frac{\mu b}{2\pi}\,\frac{x}{r^2}\Big\{1-\kappa r K_1(\kappa r)\Big\}
-\tau\mu x,
\end{align}
and in cylindrical coordinates
\begin{align}
\label{T-acyl}
\sigma_{z\varphi}=\sigma_{\varphi z}=
\frac{\mu b}{2\pi}\,\frac{1}{r}\Big\{1-\kappa r K_1(\kappa r)\Big\}-\tau\mu r.
\end{align}
We determine $\tau$ from the condition of vanishing torque,
$M_z=0$, with 
\begin{align}
M_z&=\int_0^{2\pi}\!\d\varphi\int_0^R\!\d r\,r^2 \sigma_{z\varphi}\nonumber\\
   &=\frac{\mu b R^2}{2}\Big\{1-\frac{4}{\kappa^2 R^2}+2K_2(\kappa R)\Big\}
        -\frac{1}{2}\tau\mu\pi R^4.
\end{align}
We find the ``modified'' Eshelby twist
\begin{align}
\tau=\frac{b}{\pi R^2}\Big\{1-\frac{4}{\kappa^2 R^2}+2K_2(\kappa R)\Big\}.
\end{align}
The first term is the ``classical'' Eshelby twist (see~\cite{eshelby}) 
and the last two terms are correction terms in our approach. 
The correction terms depend on the intrinsic material length $\kappa^{-1}$. 
The Eshelby twist is an observed quantity in a thin metal whisker
which has a screw dislocation along its axis~\cite{HL,twist1,twist2}.
It is important in transmission electron microscopy since it reveals 
screw dislocations normal to the foil under consideration by a 
characteristic contrast.
In principle, if the ``correction'' 
terms give a measurable contribution, the constitutive constant $\kappa$ 
could be determined in such experiment.   

Finally, the total displacements and stresses for the axial screw dislocation
in a free rod are given by
\begin{align}
&u^T_\varphi=-\frac{b r z}{\pi R^2}\Big\{1-\frac{4}{\kappa^2 R^2}+2K_2(\kappa R)\Big\},\quad
u^T_z=\frac{b}{2\pi}\,\varphi \Big\{1-\kappa r K_1(\kappa r)\Big\},\\
&\sigma_{z\varphi}=\sigma_{\varphi z}=
\frac{\mu b}{2\pi}\,\frac{1}{r}\Big\{1-\kappa r K_1(\kappa r)
-\frac{2 r^2}{R^2}\Big(1-\frac{4}{\kappa^2 R^2}+2K_2(\kappa R)\Big)\Big\}.
\end{align}
The rotation vector of the screw dislocation in a whisker reads
\begin{align}
\omega_x&=\frac{b}{4\pi}\,\frac{x}{r^2}\Big\{1-\kappa r K_1(\kappa r)\Big\}+\frac{1}{2}\, \tau x,\nonumber\\
\omega_y&=\frac{b}{4\pi}\,\frac{y}{r^2}\Big\{1-\kappa r K_1(\kappa r)\Big\}+\frac{1}{2}\, \tau y,\nonumber\\
\omega_z&=-\tau z.
\end{align}
If we use the definition of the rotation gradient (deWit's bend-twist tensor)~\cite{deWit}
\begin{align}
k_{ij}=\pd_j\omega_i,
\end{align}
the far-reaching rotation gradients of a coaxial screw dislocation in the whisker 
read
\begin{align}
k_{xx}&=\frac{b}{4\pi r^4}\Big\{\big(y^2-x^2\big)\big(1-\kappa r K_1(\kappa r)\big)
                +\kappa^2 x^2 r^2 K_0(\kappa r)\Big\}+\frac{1}{2}\,\tau,\nonumber \\
k_{yy}&=\frac{b}{4\pi r^4}\Big\{\big(x^2-y^2\big)\big(1-\kappa r K_1(\kappa r)\big)
                +\kappa^2 y^2 r^2 K_0(\kappa r)\Big\}+\frac{1}{2}\,\tau,\nonumber\\
k_{xy}&=k_{yx}=-\frac{b}{4\pi r^4}\, xy \Big\{2\big(1-\kappa r K_1(\kappa r)\big)
                -\kappa^2 r^2 K_0(\kappa r)\Big\},\nonumber\\
k_{zz}&=-\tau,\nonumber\\
k_{jj}&=\frac{1}{2}\,\alpha_{zz}.
\end{align}
For $\tau=0$, these rotation gradients are 
in agreement with the expressions calculated within the theory of Cosserat 
media (see~\cite{Kessel70,Nowacki74,Minagawa77}). 
Additionally, we find the relation $\kappa_{jj}+k_{jj}=0$. 
Thus, the sum of the Nye tensor and the rotation gradient, which may be considered 
as the total ``lattice'' curvature or total bend-twist tensor, is a deviator,
that is, it is traceless.

It is interesting to note that the strain energy of an axial screw dislocation 
is influenced from the free-surface terms.
Eventually, the strain energy of the screw dislocation in a finite cylinder of 
outer radius $R$ is given by
\begin{align}
\label{E_strain3}
 E_{\rm strain}&=\frac{\mu b^2}{4\pi} 
                \Big\{\ln\frac{\kappa R}{2}+\gamma-\frac{1}{2}
                +2K_0(\kappa R)\nonumber\\
                &\qquad\qquad\qquad
                +\frac{\kappa^2 R^2}{2}\big(K_1(\kappa R)^2
                -K_0(\kappa R)K_2(\kappa R)\big)\Big\}\nonumber\\
                &\quad -\frac{\mu b^2}{4\pi} 
                \Big\{1-\frac{8}{\kappa^2R^2}+\frac{16}{\kappa^4R^4}
                +4K_2(\kappa R)-\frac{16}{\kappa^2 R^2}\, K_2(\kappa R)\nonumber\\
                &\qquad\qquad\qquad 
                +4 K_0(\kappa R)K_2(\kappa R)
                +\frac{8}{\kappa R}\,K_1(\kappa R)K_2(\kappa R)\Big\}.
\end{align}
The terms in the third and fourth line of Eq.~(\ref{E_strain3}) arise from 
the ``modified'' Eshelby twist. 
Note that the first term in the third line of Eq.~(\ref{E_strain3})
is the correction term resulting solely from the ``classical'' Eshelby twist,
see~\cite{HL}.
The core energy (\ref{E_core}) is not influenced by the Eshelby twist.

\section{Conclusions}
In this paper we have considered a linear elastoplastic field theory of dislocations. 
The total Lagrangian of elastoplasticity has the following symbolic form
\begin{align}
\label{L-total}
{\cal L}= C\,{\boldsymbol \beta}^2+ A\,{\boldsymbol \alpha}^2
+{\text{boundary terms}},
\end{align}
where $A$, $C$ are material tensors.
It is given in terms of the both physical state quantities,
namely the elastic distortion 
${\boldsymbol \beta}$ and the dislocation density ${\boldsymbol \alpha}$.
Comparing the dimensions, we note that ${\boldsymbol \beta}$ is dimensionless
and ${\boldsymbol \alpha}$ has the dimension of an inverse length. 
Therefore, the second term in Eq.~(\ref{L-total}) contains a factor 
$\kappa^2=1/(\text{length})^2$ when compared with the first one.
In this theory,  localized 
moment stress is the specific response quantity to dislocation density.
Thus, an elastoplastic dislocation theory is a theory of force and moment stress.
Additionally, we have shown that the field equation of ${\boldsymbol \beta}$
is really the gauge theoretical modification of Kr{\"o}ner's incompatibility
equation. 

A new material coefficient $a_1$ enters
the constitutive relation between dislocation density and moment stress. 
It defines a new characteristic length $\kappa^{-1}$. 
In general, this length specifies the plastic region and
has to be estimated by means of experimental observations and 
computer simulations of the core region.
In the limit $\kappa^{-1}\rightarrow 0$, one recovers classical elasticity.

Exact analytical solutions for the total displacement field, elastic and 
plastic strain, stress fields, and  total energy of a screw dislocation
in a cylinder and of a coaxial screw dislocation
have been reported which have not any singularity in the dislocation
core region. Consequently, it would be interesting to compare them with 
experimental results and related computer simulations.
For the coaxial screw dislocation, we found a modified formula for the 
so-called Eshelby twist. The results for the screw dislocation in a thin whisker
should be useful for torsion experiments of thin wires.

Note that the present theory of elastoplasticity is intimately 
connected with the so-called gradient theory of dislocations~\cite{GA96,GA99}
and Eringen's nonlocal elasticity~\cite{eringen83,eringen85,eringen87,eringen90}. 
The stress field of a screw dislocation in elastoplasticity coincides
with the solution in Eringen's nonlocal 
elasticity~\cite{eringen83,eringen85,eringen87,eringen90}
and in gradient elasticity~\cite{GA99,Gutkin00}.
The internal characteristic length is thus connected to the 
gradient coefficient and to the nonlocality parameter.
Obviously, the equations (\ref{fe-E}) and (\ref{stress-fe-screw})
have the form of gradient theory equations of second order. 
The new material coefficient $a_1$ 
has the meaning similar as a gradient coefficient.

\subsection*{Acknowledgement}
The author would like to thank the Graduate College 
``Quantum field theory'' at the Center for Theoretical Studies of Leipzig
University for financial support.  
He acknowledges the support provided by the Max-Planck Institute
for Mathematics in the Sciences in Leipzig during the final stage 
of this work.

\end{document}